\documentclass[10pt]{article}
\usepackage{graphics} % for pdflatex is that images must 
\usepackage{array,times,url}
\usepackage{epsfig}

\newtheorem{theorem}{Theorem}[section]
\newtheorem{lemma}[theorem]{Lemma}

\newtheorem{corollary}[theorem]{Corollary}
\newtheorem{example}{Example}
\newtheorem{definition}[theorem]{Definition}

\begin{document}

\makeatletter

\title{Homogeneous Faults, Colored Edge Graphs, and Cover Free Families}
\author{Yongge Wang, UNC Charlotte, USA, yonwang@uncc.edu\\
Yvo Desmedt, University College London, UK}
\maketitle

\thispagestyle{empty}
\begin{abstract}
In this paper, we use the concept of colored edge
graphs to model homogeneous faults in networks. We then use
this model to study the minimum connectivity (and design)
requirements of networks for being robust against 
homogeneous faults within certain thresholds.
In particular, necessary and sufficient conditions for 
most interesting cases are obtained. For example, we will study the
following cases: (1) the number of colors (or the number of non-homogeneous 
network device types) is one more than the homogeneous fault threshold;
(2) there is only one homogeneous fault (i.e., only one color could fail);
and (3) the number of non-homogeneous 
network device types is less than five.
\end{abstract}

\setcounter{page}{1}
\section{Background and colored edge graph}
In network communications, the communication could fail
if some nodes or some edges are broken. 
Though the failure of a modem could be considered the failure of 
a node, we can model this scenario also as the failure of the 
communication link (the edge) attached to this modem. 
Thus it is sufficient to consider edge failures in communication networks.
It is also important to note that several nodes (or edges) in a network
could fail at the same time. For example, all brand X routers
in a network could fail at the same time due to a 
platform dependent computer worm (virus) attack.
In order to design survivable communication networks, it is essential
to consider this kind of homogeneous faults for networks. Existing 
works on network quality of services have not addressed this 
issue in detail and there is no existing model to study network reliability 
in this aspect. In this paper,
we use the colored edge graphs which could be used to 
model homogeneous faults in networks. The model is then used
to optimize the design of survivable networks
and to study the minimum connectivity (and design)
requirements of networks for being robust against 
homogeneous faults within certain thresholds.

\begin{definition}
A colored edge graph is a tuple $G(V,E,C,f)$, with $V$ the node set,
$E$ the edge set, $C$ the color set, and $f$ a map from $E$ onto $C$.
The structure
$$
{\cal Z}_{C,t}=\{Z: Z\subseteq E\,\, \hbox{and}\,\,\,
|f(Z)|\leq t\}.
$$
is called a $t$-{\it color\/} adversary structure.
Let $A,B \in V$ be distinct nodes of $G$.
$A,B$ are called $(t+1)$-{\em color connected\/} for $t\ge 1$ if 
for any color set $C_{t}\subseteq C$ of size $t$, 
there is a path $p$ from $A$ to $B$ 
in $G$ such that the edges on $p$ do not contain any color in $C_{t}$.
A colored edge graph $G$ is  $(t+1)$-{\em color connected} if and only
if for any two nodes $A$ and $B$ in $G$, they are 
$(t+1)$-color connected.
\end{definition}

The interpretation of the above definition is as follows. In a network,
if two edges have the same color, then they could fail at the same
time. This may happen when the two edges are designed with same
technologies (e.g., with same operating systems, with same 
application software, with same hardware, or with same 
hardware and software). If a colored edge network is 
$(t+1)$-color connected, then the network communication
is robust againt the failure of edges of any $t$ colors
(that is, the adversary may tear down any $t$ types of devices).

In practice, one communication link may be attached to different brands of 
network devices (e.g., routers, modems) on both sides. 
For this case, the edge can have two different colors.
If any of these colors is broken, the edge is broken. 
Thus from a reliability viewpoint, if one designs networks with  
two colors on the same edge, the same reliability/security can be 
obtained by having only one color on each edge. In the following 
discussion, we will only consider the case with one color
on each edge. Meanwhile, multiple edges between two nodes are 
not allowed either.

We are interested in the following practical questions. For a given  
number $n$ of nodes in $V$ (i.e., the number of network nodes), 
a given number $m$ of the colors (e.g., the number of network device types), 
and a given number $t$, how can we design a $(t+1)$-color connected
colored edge graphs $G(V,E)$ with minimum 
number $\lambda$ of edges? In another words, how can we use minimum
resources (e.g., communication links) to design a network
that will keep working even if $t$ types of devices
in the network fail?

For practical network designs, one needs first to have an estimate
on the number of homogeneous faults. For example, the number $t$ of 
brands of routers that could fail at the same time. Then it is sufficient
to design a $(t+1)$-color connected network with $m=t+1$ colors (e.g., with 
$t+1$ different brands of routers). Necessary and sufficient conditions
for this kind of network design will be obtained in this paper.

Another important issue that should be taken into consideration 
in practical network designs is that the number
$m$ of colors (e.g., the number of brands for routers) is 
quite small. For example, $m$ is normally less than five.
Necessary and sufficient conditions for network designs with
$m\le 5$ and with optimized resources 
will be obtained in this paper. Note that for cases with small
$m$, we may have $m>t+1$.

The outline of the paper is as follows. Section \ref{meqtp1sec}
describes the necessary and sufficient conditions for the case 
of $m=t+1$ without optimizing the number of edges in the networks. 
Section \ref{generalnecsec} gives a necessary condition for colored
edge networks in terms of optimized number of edges.
Section \ref{practiffsec} shows that the necessary conditions in 
Section \ref{generalnecsec} are also sufficient for the most 
important three cases: (1) $m=t+1$; (2) $t=1$; and (3) $m\le 5$. 
Section \ref{hardsec} shows that it is {\bf NP}-hard to determine whether
a given colored edge graph is $(t+1)$-connected.

\section{Related works}
Though colored-edge graph is a new concept which we used
to model network survivability issues, there are 
related research topics in this field. For example,
edge-disjoint (colorful) spanning trees have been extensively
studied in the literature (see, e.g., \cite{constantine}).
These results are mainly related to our discussion in the next
section for the case of $m=t+1$. 
A colored edge graph $G$ is {\em proper} if whenever 
two edges share an end point they carry different colors.
A spanning tree for a colored edge graph is called 
colorful if no two of its edges have the same color.
Two spanning trees of a graph are edge disjoint 
if they do not share common edges. For a non-negative
integer $s$, let $K_{s}$ denote the complete graph on $s$ vertices.
A classical result of Euler states that the edges of $K_{2n}$
can be partitioned into $n$ isomorphic spanning trees (paths, for example)
and each of these spanning trees can easily be made colorful, but 
the resulting edge colored graph usually fails to be proper.

Though it is important to design colored edge graphs with
required security parameters, for several scenarios 
it is also important to calculate the robustness of 
a given colored edge graphs. Roskind and Tarjan \cite{roskind} designed 
a greedy algorithm to find $(t+1)$-edge disjoint spanning 
trees in a given graph. This is related to the 
questions $(t+1)$-color connectivity for the case of $m=t+1$.
We are not aware of any approximate algorithms 
for deciding $(t+1)$-color connectivity of a given colored edge graph.
Indeed, we will show that this problem is {\bf NP}-hard.

\section{Necessary and sufficient conditions for $m=t+1$}
\label{meqtp1sec}
In this section, we show necessary and sufficient conditions for 
some special cases.
\begin{lemma}
\label{lemma1s}
A colored edge graphs $G(V,E,C,f)$ is 
$(t+1)$-color connected if and only if, for all $i_1$, $i_2$,
$\ldots$, $i_{m-t}\le m$,
$(V, E_{i_1}\cup E_{i_2}\cup\cdots\cup E_{i_{m-t}})$ is 
a connected graph, where 
$E_1,E_2,\ldots,E_m$ is a partition of $E$ under the $m$ different colors.
\end{lemma}

As we have mentioned in the previous section, the classical
result by Euler states that $K_{2n}$ can be partitioned 
into $n$ spanning trees. Thus, by Lemma \ref{lemma1s}, 
we have the following theorem.

\begin{theorem}
\label{eulerresult}
(Euler) For $n=2m$, there is a coloration $G(V,E,C,f)$ of $K_n$
such that $G$ is $(m-1)$-color connected.
\end{theorem}

In the following, we extend  Theorem \ref{eulerresult} to the
general case of $n\ge 2m$.

\begin{lemma}
\label{lemma2s}
For $n\ge 2m$ and $m\ge 2$, there exists a graph $G(V,E)$ with 
$|V|=n, |E|=m(n-1)$, and $E=E_1\cup E_2
\cup \cdots\cup E_m$ such that the following conditions are satisfied:
\begin{enumerate}
\item $G(V,E_i)$ is a connected graph for all $0<i\le m$;
\item $E_i\cap E_j=\emptyset$ for all $i,j\le m$.
\end{enumerate}
\end{lemma}

\noindent
{\bf Proof.} We prove the Lemma by induction on $n$ and $m$. 
For $n=2$ and $m=1$,
the Lemma holds obviously.
Assume that the Lemma holds for $n_0=2m_0$. 

In the following, 
we show that the Lemma holds
for $n=n_0+1, m=m_0$ and for $n=n_0+2, m=m_0+1$.
Let $G(V_0,E_0)$ be the graph with 
$|V_0|=n_0, |E_0|=m_0(n_0-1)$, and $E_0=E^0_1\cup E^0_2
\cup \cdots\cup E^0_{m_0}$ such that the conditions in 
the Lemma are satisfied:

For the case of  $n=n_0+1$ and $m=m_0$, let $V=V_0\cup \{u\}$
where $u$ is a new node that is not in $V_0$, and let 
$E_1=E^0_1\cup \{(u,u_1)\}$, $E_2=E^0_2\cup \{(u,u_2)\}$, 
$\ldots$, $E_{m_0}=E^0_{m_0}\cup \{(u,u_{m_0})\}$ where
$u_1, u_2, \ldots, u_{m_0}$ are distinct nodes from $V_0$.
It is straightforward to show that $|V|=n, |E|=m(n-1)$,
$G(V,E_i)$ is a connected graph, and 
$E_i\cap E_j=\emptyset$ for all $i,j\le m$. Thus 
the Lemma holds for this case.

For the case of  $n=n_0+2$ and $m=m_0+1$, let $V=V_0\cup \{u,v\}$
where $u,v$ are new nodes that are not in $V_0$, and 
define $E_1, \ldots, E_m$ as follows.
\begin{enumerate}
\item Set $E_m=\emptyset$ and $U=\emptyset$, 
where $U$ is a temporary variable.
\item Define $E_1$:
\begin{enumerate}
\item Select an edge $(v_1,v_2)\in E^0_1$.
\item Let 
$E_1=\left(E_1^0\setminus \{(v_1,v_2)\}\right)\bigcup 
   \{(v_1,u), (u,v), (v,v_2)\}$.
\item Let $E_m=E_m\cup\{(v,v_1), (v_1,v_2), (v_2,u)\}$ and 
$U=U\cup\{v_1,v_2\}$.
\end{enumerate}
\item Define $E_i$ for $2\le i\le m_0$:
\begin{enumerate}
\item Select $v_{2i-1},v_{2i}\notin U$. 
\item Let $E_i=E_i^0\cup\{(u,v_{2i-1}), (v,v_{2i})\}$.
\item Let $E_m=E_m\cup\{(v,v_{2i-1}), (u,v_{2i})\}$ and 
$U=U\cup\{v_{2i-1},v_{2i}\}$.
\end{enumerate}
\end{enumerate}
It is straightforward to show that $|V|=n, |E_i|=(n-1)$ 
(thus $|E|=m(n-1)$),
$G(V,E_i)$ is a connected graph, and 
$E_i\cap E_j=\emptyset$ for all $i,j\le m$. This completes
the proof of the Lemma.
\hfill{Q.E.D.}

\begin{theorem}
\label{iffthmt1}
Given $n, m, t$ with $m=t+1$, there exists a 
$(t+1)$-color connected colored edge graphs $G(V,E,C,f)$ 
with $|V|=n$ and $|C|=m$ if and only if $n\ge 2m$.
\end{theorem}

\noindent
{\bf Proof.}  By Lemma \ref{lemma1s}, a $(t+1)$-color connected 
colored edge graphs $G(V,E,C,f)$ with $|V|=n$ and $|C|=m=t+1$ 
contains at least $m(n-1)$ edges. Meanwhile, $G(V,E,C,f)$
contains at most $n(n-1)/2$ edges. Thus for $n< 2m$, 
we have  $n(n-1)/2<m(n-1)$. In nother words, for $n< 2m$,
there is no $(t+1)$-color connected colored edge graphs $G(V,E,C,f)$ 
with $|V|=n$ and $|C|=m=t+1$.  Now the theorem follows from 
Lemmas \ref{lemma1s} and \ref{lemma2s}.
\hfill{Q.E.D}

\section{Necessary conditions for general cases}
\label{generalnecsec}
First we note that for a colored edge graph $G$ to be $(t+1)$-color connected,
each node must have a degree of at least $t+1$. Thus the total degree
of an $n$-node graph should be at least $n(t+1)$. This implies the following
lemma.

\begin{lemma}
\label{nesslemma1}
For $m\ge t+1>1$, and a $(t+1)$-color connected colored edge graph 
$G(V,E,C,f)$ with $|V|=n$, $|E|=\lambda$, and $|C|=m$, we 
have $2\lambda\ge (t+1)n$.
\end{lemma}

In the following, we use cover free family concepts to study the necessary
conditions for colored edge graphs connectivity.

\begin{definition}
Let $X$ be a finite set with $|X|=\lambda$ and
${\cal F}$ be a set of mutually disjoint subsets of $X$
with $|{\cal F}|=m$. Then $(X,{\cal F})$ is called a $(\lambda,m)$-partition 
of $X$ if $X=\bigcup_{P\in {\cal F}}P$. 
Let $n,t$ be positive integers. An $(\lambda,m)$-partition $(X,{\cal F})$ 
is called a $(t;n-1)$-cover free family (or $(t;n-1)$-CFF($\lambda,m$)) 
if, for any $t$ elements $B_1, \ldots, B_t\in {\cal F}$, 
we have that
$$\left|X\setminus \left(\bigcup_{i=1}^{t}B_i\right)\right|\ge n-1
\quad\quad\left(\mathrm{or }
\left|\bigcap_{i=1}^{t}
\left(X\setminus B_i\right)\right|\ge n-1
\right)$$
\end{definition}

It should be noted that our above definition of cover-free
family is different from the generalized cover-free 
family definition for set systems in the literature 
(see, e.g., \cite{engel,stinsonwei}).
In \cite{stinsonwei}, a set system $(X,{\cal F})$ is called a 
$(w,t;n-1)$-cover free family if for 
any $w$ blocks $A_1, \ldots, A_w\in {\cal F}$ and
any $t$ blocks $B_1, \ldots, B_t\in {\cal F}$, one has 
$\left|\left(\cap_{j=1}^w A_j\right)\setminus 
\left(\cup_{i=1}^{t}B_i\right)\right|\ge n-1$.
Specifically, there are two major differences between
our $(\lambda, m)$-partition system and the set systems in the 
literature\footnote{The first author of this paper would like to thank 
Prof. Doug Stinson for pointing this out to the author.}.
\begin{enumerate}
\item For a set system $(X,{\cal F})$, ${\cal F}$ may contain repeated
elements.
\item For a set system $(X,{\cal F})$, the elements in ${\cal F}$ 
are not necessarily mutually disjoint.
\end{enumerate}

It is straightforward to show that a colored edge graph $G$ is 
$(t+1)$-color connected if and only if for any color set 
$C_{t}\subseteq C$ of size $t$, after the removal of edges
in $G$ with colors in $C_t$, $G$ remains connected. Assume that
$G$ contains $n$ nodes. Then a necessary condition for connectivity
is that $G$ contains at least $n-1$ edges. 
From this discussion, we get the following lemma.

\begin{lemma}
For a colored edge graph $G(V,E,C,f)$, with $|V|=n$, $|E|=\lambda$,
$|C|=m$, a necessary condition for  $G(V,E,C,f)$ to be 
$(t+1)$-color connected is that the $(\lambda,m)$-partition
$(X,{\cal F})$ is a $(t;n-1)$-CFF($\lambda, m$) with
$X=E$ and ${\cal F}=\{E_c: c\in C\}$ where 
$E_c=\{e: f(e)=c, e\in E\}$.
\end{lemma}

In the following, we analyze lower bounds for the number $\lambda$ 
of edges for the existence of a $(t;n-1)$-CFF($\lambda, m$).
For a set partition $(X,{\cal F})$ and a positive integer $t$, let 
$$\mu{(X,{\cal F};t)}=\min\left\{
\left|X\setminus \left(\bigcup_{i=1}^{t}B_i\right)\right|:
{B_1,\ldots, B_t\in {\cal F}}\right\}$$
It is straightforward to see that a $(\lambda,m)$-partition $(X,{\cal F})$ is 
a $(t;n-1)$-CFF($\lambda, m$) if and only if $\mu{(X,{\cal F};t)}\ge n-1$.

Given positive integers $\lambda, m,t$, let 
$$\mu(\lambda,m;t)=
\max\left\{\mu{(X,{\cal F};t)}:(X,{\cal F}) 
\mbox{ is a }{(\lambda,m)\mbox{-partition}}\right\}$$
From the above discussion and Lemma \ref{nesslemma1}, 
we have the following theorem.
\begin{theorem}
\label{dogthm}
Let $\lambda, m,t$ be given positive integers. 
$\mu(\lambda,m;t)\ge n-1$ and $2\lambda\ge (t+1)n$
are necessary conditions for the 
existence of a $(t+1)$-color connected colored edge graph 
$G(V,E,C,f)$, with $|V|=n$, $|E|=\lambda$, $|C|=m$.
\end{theorem}

\begin{theorem}
\label{sufbound}
Let $\lambda, m,t$ be given positive integers. Then we have 
$$\mu(\lambda,m;t)=\left\{\begin{array}{ll}
(m-t)\cdot\lfloor\frac{\lambda}{m}\rfloor& 
         \mbox{if } t\ge \lambda-\lfloor\frac{\lambda}{m}\rfloor\cdot m\\
(m-t)\cdot\lfloor\frac{\lambda}{m}\rfloor+
\left(\lambda-\lfloor\frac{\lambda}{m}\rfloor\cdot m-t\right)
&\mbox{otherwise}
\end{array}\right.$$
\end{theorem}

\noindent
{\bf Proof.} 
For a given $(\lambda,m)$-partition $(X,{\cal F})$,
let $B_1, \ldots, B_m$ be an enumeration of elements in ${\cal F}$
such that $|B_i|\le |B_{i+1}|$ for all $i<m$.
It is straightforward to show that 
$\mu{(X,{\cal F};t)}=\sum_{i=1}^{m-t}|B_i|$. Thus $\mu(\lambda,m;t)$ takes the 
maximum value if $\sum_{i=1}^{m-t}|B_i|$ is maximized. It is 
straightforward to show that this value is maximized when 
the $(\lambda,m)$-partition $(X,{\cal F})$ satisfies the following conditions:
\begin{enumerate}
\item $|B_i|=\lfloor\frac{\lambda}{m}\rfloor$
for $i\le m-\left(\lambda-\lfloor\frac{\lambda}{m}\rfloor\cdot m\right)$, and 
\item $|B_i|=\lfloor\frac{\lambda}{m}\rfloor+1$ for 
$m\ge i> m-\left(\lambda-\lfloor\frac{\lambda}{m}\rfloor\cdot m\right)$.
\end{enumerate}
The theorem follows from the above discussion.
\hfill{Q.E.D.}

\begin{example}
For $n=7, \lambda=10, m=5$, and $t=2$, we have 
$\mu(10,5;2)=6= n-1$. However, $2\lambda = 20 < (t+1)n=21$.
This shows that the condition $2\lambda\ge (t+1)n$
in Theorem \ref{dogthm} is not redundant.
\end{example}

\begin{example}
There are no $(t+1)$-color connected colored edge graphs $G(V,E,C,f)$ 
for the following special cases:
\begin{enumerate}
\item $m=2,t=1, n=3$.
\item $m=4,t=2,n=4$.
\item $m=3,t=2, n\le 5$.
\end{enumerate}
\end{example}

\noindent
{\bf Proof.} Before we consider the specific cases, we observe that, 
when $m$ and $t$ are fixed, the function $\mu$ is nondecreasing 
when $\lambda$ increases.

1. In this case, the maximum value that $\lambda$ could 
take is $3$. Thus $\mu(3,2;1)=1<n-1=2$. 
That is, there is no $(1;2)$-CFF($3,2$), which implies the claim.
Note that this result also follows from Theorem \ref{iffthmt1}.

2. In this case, the maximum value that $\lambda$ could 
take is $6$. Thus $\mu(6,4;2)=2<n-1=3$. 

3. We only show this for the case $m=3,t=2, n=5$. In this case,
the maximum value that $\lambda$ could 
take is $10$. Thus $\mu(10,3;2)=1<n-1=4$.
Note that this result also follows from Theorem \ref{iffthmt1}.
\hfill{Q.E.D}

\vskip 10pt
The following theorem is a variant of Theorem \ref{dogthm}.

\begin{theorem}
\label{necessarybound}
For $m-1>t>0$, a necessary condition for the existence of 
a $(t+1)$-color connected colored edge graph 
$G(V,E,C,f)$ with $|V|=n$, $|E|=\lambda$, and $|C|=m$ is that 
$2\lambda\ge (t+1)n$ and 
the following conditions are satisfied:
\begin{itemize}
\item If $n=(m-t)k$\/ for some integer $k>0$, then $\lambda \ge mk-1$.
\item If $n=(m-t)k+1$\/ for some integer $k>0$, then $\lambda \ge mk$.
\item If $n=(m-t)k+2$\/ for some integer $k>0$, then $\lambda \ge mk+t+1$.
\item $\cdots\cdots$
\item If $n=(m-t)k+m-t-1$\/ for some integer $k>0$, then $\lambda \ge mk+m-2$.
\end{itemize}
\end{theorem}

\noindent
{\bf Proof.}
For $m>t+1$, by Theorem \ref{sufbound}, we have 
$$\mu(\lambda,m;t)=\left\{\begin{array}{ll}
(m-t)k'&\mbox{ if }\lambda = mk'+i \mbox{ for }\ 0\le i\le t \\
(m-t)k'+1&\mbox{ if } \lambda = mk'+t+1\\
\cdots\cdots\\
(m-t)k'+m-t-1&\mbox{ if } \lambda = mk'+m-1
\end{array}\right.$$
Thus the necessary condition $\mu(\lambda,m;t)\ge n-1$ in 
Theorem \ref{dogthm} can be interpreted as the following conditions: 
\begingroup
\def\arraystretch{1.3}
$$k'\ge\left\{\begin{array}{ll}
\frac{n-1}{m-t}&\mbox{ if }\lambda = mk'+i \mbox{ for }\ 0\le i\le t \\
\frac{n-2}{m-t}&\mbox{ if } \lambda = mk'+t+1\\
\cdots\cdots\\
\frac{n-m+t}{m-t}&\mbox{ if } \lambda = mk'+m-1
\end{array}\right.$$
\endgroup
In other words, for a $(t+1)$-color connected colored edge
graphs $G(V,E,C,f)$, the following $m-t$ conditions
(the disjunction not conjunction) are satisfied:
\begin{itemize}
\item $|V|=n, |E|\ge m\left\lceil\frac{n-1}{m-t}\right\rceil$, and $|C|=m$.
\item $|V|=n, |E|\ge m\left\lceil\frac{n-2}{m-t}\right\rceil+t+1$, and $|C|=m$.
\item $\cdots\cdots$
\item $|V|=n, |E|\ge m\left\lceil\frac{n-m+t}{m-t}\right\rceil+m-1$, 
   and $|C|=m$.
\end{itemize}
By distinguishing the cases for $n=(m-t)k$, $n=(m-t)k+1$,
$\cdots$, and $n=(m-t)k+m-t-1$, and by reorganizing above lines,
these necessary conditions can be interpreted as the 
following $m-t$ conditions:
\begin{itemize}
\item $n=(m-t)k$ and $\lambda \ge mk-1$ for some $k>0$. Note that
this follows from the last line of the above conditions (one can surely
take other lines, but then the value of $\lambda$ would be larger).
This comment applies to following cases also.
\item $n=(m-t)k+1$ and $\lambda \ge mk$ for some $k>0$.
\item $n=(m-t)k+2$ and $\lambda \ge mk+t+1$ for some $k>0$.
\item $\cdots\cdots$
\item $n=(m-t)k+m-t-1$ and $\lambda \ge mk+m-2$ for some $k>0$.
\end{itemize}
\hfill{Q.E.D.}

\section{Necessary and sufficient conditions for practical cases (with 
small $m$ and $t$)}
\label{practiffsec}
Generally we are interested in the question whether the necessary condition 
in Theorems \ref{dogthm} and \ref{necessarybound} are also sufficient. 
In the following, we show that this is true for several important 
practical cases.

\begin{theorem}
\label{mistp1}
The necessary condition in Theorem \ref{dogthm} is sufficient
for the case of $m=t+1$.
\end{theorem}

\noindent
{\bf Proof.}
Since $\lambda-\lfloor\frac{\lambda}{m}\rfloor\cdot m$ is the remainder of
$\lambda$ divided by $m$, we trivially have $t=m-1\geq 
\lambda-\lfloor\frac{\lambda}{m}\rfloor\cdot m$.
Now assume that $m>\frac{n}{2}$. By Theorem \ref{sufbound}, we 
have $\mu(\lambda,m;t)=\lfloor\frac{\lambda}{m}\rfloor\le 
\lfloor\frac{n(n-1)}{2m}\rfloor<n-1$. 
The rest follows from Theorem \ref{iffthmt1}.
\hfill{Q.E.D.}

\vskip 10pt
Before we show that the necessary conditions 
in Theorems \ref{dogthm} and \ref{necessarybound} are 
sufficient for the case of $t=1$,
we first present two lemmas whose proofs are straightforward.

\begin{lemma}
For $n=m=\lambda\ge 3$ and $t=1$, the following $m$-node circle graph
is $(1+1)$-color connected: 
$$\left\{(v_1,v_2), (v_2, v_3), \ldots, (v_{m},v_1)\right\}$$ 
with $f(v_i,v_{i+1})=c_{i}$ for $i<m$ and $f(v_m,v_{1})=c_{m}$.
\end{lemma}

\begin{lemma}
\label{lemmacircle}
For $t=1$, $m\ge 3$, and $m< n\le 2m-2$, the graph in Figure 
\ref{figcircle} that is defined in the following is $(1+1)$-color connected 
$$\left\{(v_1,v_2), (v_2, v_3), \ldots, (v_{m},v_1)\right\}\cup
\left\{(v_m, v_{m+1}), (v_{m+1},v_{m+2}), \ldots, (v_n,v_1)\right\}$$
with 
$$\begin{array}{rll}
f(v_i,v_{i+1})&=c_{i} & \mbox{ for } 1\le i\le m-1\\
f(v_{m},v_1)&=c_m &\\
f(v_{m+i-1},v_{m+i})&=c_i &\mbox{ for }1\le i\le n-m\\
f((v_n,v_1))&=c_{n-m+1}&
\end{array}$$
\begin{figure}[htb]
\centering{\includegraphics{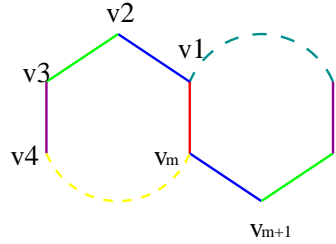}}
%\centering{\includegraphics[height=3in, width=4in]{circle}}
\caption{Graph for Lemma \ref{lemmacircle}}
\label{figcircle}
\end{figure}
\end{lemma}

\begin{theorem}
\label{tp1case}
The necessary conditions in Theorems \ref{dogthm} and \ref{necessarybound}
are sufficient for the case of $t=1$.
\end{theorem}

\begin{figure}[htb]
\centering{\includegraphics{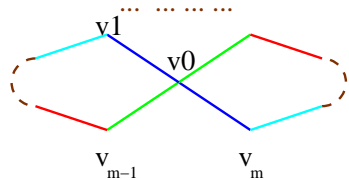}}
%\centering{\includegraphics[height=3in, width=4in]{t1}}
\caption{Graph for the case $n=(m-1)k+1$ and $\lambda \ge km$}
\label{figt1basic}
\end{figure}

\noindent
{\bf Proof.}
For the case of $m=2$ and $t=1$, it follows from Theorem \ref{mistp1}.
Now assume that $m>2$ and $t=1$. In this special case, 
the necessary conditions in Theorem \ref{necessarybound} is as follows:
\begin{itemize}
\item $n=(m-1)k$ and $\lambda \ge mk-1$ for some $k>0$.
\item $n=(m-1)k+1$ and $\lambda \ge mk$ for some $k>0$.
\item $n=(m-1)k+2$ and $\lambda \ge mk+2$ for some $k>0$.
\item $\cdots\cdots$
\item $n=(m-1)k+m-2$ and $\lambda \ge mk+m-2$ for some $k>0$.
\end{itemize}
In the following we first show that the condition
``$n=(m-1)k+1$ and $\lambda \ge km$'' is sufficient. Let the graph 
in Figure \ref{figt1basic} 
be defined as follows:
$$\begin{array}{lll}
V&=&\{v_0, v_1\cdots, v_{(m-1)k}\},\\ 
E_1&=&\{(v_0, v_{(m-1)i+1}): 0\le i\le k-1\}\\
E_j&=&\{(v_{(m-1)i+j-1}, v_{(m-1)i+j}): 
    0\le i\le k-1\}\mbox{ for } 2\le j\le m-1\\
E_m&=&\{(v_{(m-1)i},v_0): 1\le i\le k\}\\
E&=&E_1\cup E_2\cup \cdots \cup E_m
\end{array}$$
For each $e\in E_j$ with $i\le m$, let $f(e)=c_j$. Then it 
is straightforward to check that the colored edge graphs 
$G(V,E,C,f)$ is $(1+1)$-color connected, $|V|=(m-1)k+1$, and $|E|=mk$.

Now we show that the condition 
``$n=(m-1)k+j$ and $\lambda \ge km+j$ for $2\le j\le m-1$'' 
is sufficient. Let $G(V,E,C,f)$ be the colored edge
graph that we have just constructed with $|V|=(m-1)k+1$, and $|E|=mk$.

\begin{figure}[htb]
\centering{\includegraphics{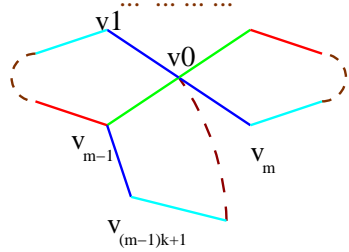}}
%\centering{\includegraphics[height=3in, width=4in]{t11}}
\caption{Graph for case $n=(m-1)k+j$ and $\lambda \ge km+j$ for $2\le j\le m-1$}
\label{figt1case1}
\end{figure}

Let $V'=V\cup \{v_{(m-1)k+1}, \ldots, v_{(m-1)k+j-1}\}$. 
Define a new colored edge graph  $G(V',E',C,f')$ (see Figure \ref{figt1case1})
by attaching the following edges to the $m$-node circle
$\{(v_0,v_1), (v_1,v_2), \ldots, (v_{m-1},v_0)\}$:
$$\{(v_{m-1},v_{(m-1)k+1}), (v_{(m-1)k+1},v_{(m-1)k+2}), \ldots,
(v_{(m-1)k+j-1},v_0)\}$$
The colors for the new edges are defined by letting
$f'(v_{(m-1)k+i},v_{(m-1)k+i+1})=c_{i+1}$ for $0\le i\le j-2$ and
$f'(v_{(m-1)k+j-1},v_0)=c_j$. It is straightforward to check that 
$G(V',E',C,f')$ is $(1+1)$-color connected, $|V|=(m-1)k+j$, and $|E|=mk+j$.
\hfill{Q.E.D.}

\begin{corollary}
For $t=1$ and $m$, $n,\lambda>1$, there exists an $(1+1)$-color 
connected colored edge graph $G(V,E,C,f)$ with $|V|=n$ and $|E|=\lambda$
if and only if 
$$\lambda\ge \min\left\{
m\left\lceil\frac{n-1}{m-1}\right\rceil, 
m\left\lceil\frac{n-2}{m-1}\right\rceil+2,
\ldots,
m\left\lceil\frac{n-m+1}{m-1}\right\rceil+m-1
\right\}.$$
\end{corollary}

\noindent
{\bf Proof.}
It follows from the proof of Theorem \ref{tp1case}.
\hfill{Q.E.D}

\begin{theorem}
\label{m4t2}
The conditions in Theorems \ref{dogthm} and \ref{necessarybound}
are sufficient for the case of  $m=4,t=2$.
\end{theorem}

\noindent
{\bf Proof.}
It is sufficient to show that 
both of the conditions ``$n=(m-t)k+1$ and $\lambda \ge km$''
and ``$n=(m-t)k+2$ and $\lambda \ge mk+t+1$'' are sufficient 
(note that $m=4$ and $t=2$).
In the following we first show that the condition
``$n=(m-t)k+1$ and $\lambda \ge km$'' is sufficient by induction
on $k$. 

\begin{figure}[htb]
\centering{\includegraphics{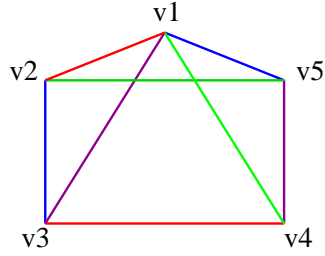}}
%\centering{\includegraphics[height=3in, width=4in]{t11.eps}}
\caption{Graph $G_1$ for the case $n=5,m=4, t=2$}
\label{figm4t2}
\end{figure}

For the case of $k=2$, we have $n=5,\lambda=8,m=4$, and $t=2$.
Let the graph $G_1$ in Figure \ref{figm4t2} be defined as
$$G_1=\{(v_1,v_2)_1, (v_2,v_3)_2,(v_3,v_4)_1,(v_4,v_5)_3,(v_5,v_1)_2,
(v_1,v_3)_3, (v_1,v_4)_4,(v_2,v_5)_4\}$$
where $(v,v')_i$ means that the edge $(v,v')$ takes color $c_i$.
It is straightforward to check that $G_1$ is $(2+1)$-color connected.

\begin{figure}[htb]
\centering{\includegraphics{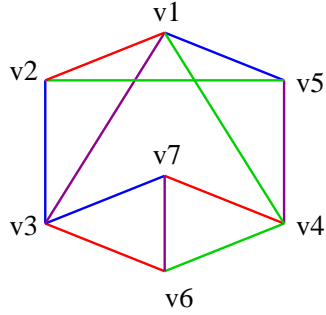}}
%\centering{\includegraphics[height=3in, width=4in]{t11.eps}}
\caption{Graph $G_2$ for the case $n=7,m=4, t=2$}
\label{fign7m4t2}
\end{figure}

For the case of $k=3$, we have $n=7,\lambda=12,m=4$, and $t=2$.
Let the graph $G_2$ in Figure \ref{fign7m4t2} be defined as 
$$\begin{array}{l}
\{(v_1,v_2)_1, (v_2,v_3)_2,(v_4,v_5)_3,(v_5,v_1)_2,
(v_1,v_3)_3, (v_1,v_4)_4,\\
\quad\quad (v_2,v_5)_4, (v_3,v_6)_1,(v_6,v_7)_3,
(v_7,v_4)_1,(v_4,v_6)_4,(v_3,v_7)_2\}
\end{array}$$
where $(v,v')_i$ means that the edge $(v,v')$ takes color $c_i$.
It is straightforward to check that $G_2$ is $(2+1)$-color connected.

Now for $k=2r$ ($r\ge 2$), we have $n=(m-t)k+1=4r+1$ and $\lambda =km=8r$. 
If we glue the $v_1$ node of $r$ copies of $G_1$, we
get a $(t+1)$-color connected colored graph $G$ with
$n=4r+1$ and $\lambda =8r$. Thus the condition for the case
of $k=2r$ holds.

For $k=2r+1$ ($r\ge 2$), we have 
$n=(m-t)k+1=4r+3$ and $\lambda=km=8r+4$.
If we glue glue the $v_1$ node of $r-1$ copies of 
$G_1$ and one copy of $G_2$, we get a $(t+1)$-color connected 
colored graph $G$ with $n=4(r-1)+1+6=4r+3$ and $\lambda =8(r-1)+12=8r+4$. 
Thus the condition for the case of $k=2r+1$ holds.
This completes the induction.

For the condition 
``$n=(m-t)k+2$ and $\lambda \ge mk+t+1$'', one can add one 
node to the graph for the case ``$n=(m-t)k+1$ and $\lambda \ge km$''
 with $3$ edges (with distinct colors) 
to any three nodes. The resulting graph meets the requirements.
\hfill{Q.E.D.}

Theorem \ref{m4t2} could be extended to the case of $m=5$
and $t=3$.
\begin{theorem}
\label{met2}
The conditions in Theorems \ref{dogthm} and \ref{necessarybound}
are sufficient for the case of $m=5$ and $t=3$.
\end{theorem}

\noindent
{\bf Proof.}
It is sufficient to show that 
both of the conditions ``$n=(m-t)k+1$ and $\lambda \ge km$''
and ``$n=(m-t)k+2$ and $\lambda \ge mk+t+1$'' are sufficient 
(note that $m-t=2$).
In the following we first show that the condition
``$n=2k+1$ and $\lambda \ge km$'' is sufficient by induction
on $k$ and $m$. 

\begin{figure}[htb]
\centering{\includegraphics{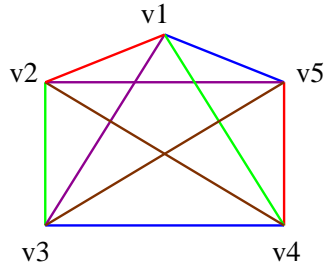}}
%\centering{\includegraphics[height=3in, width=4in]{t11.eps}}
\caption{Graph $G_{5,1}$ for the case $n=5,m=5, t=3$}
\label{fign5m5t3}
\end{figure}

\begin{figure}[htb]
\centering{\includegraphics{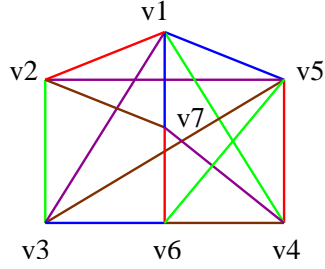}}
%\centering{\includegraphics[height=3in, width=4in]{t11.eps}}
\caption{Graph  $G_{5,2}$ for the case $n=7,m=5, t=3$}
\label{fign7m5t3}
\end{figure}

For $m=5$ and $k=2$, we have $n=5, \lambda=10$. The graph in Figure
\ref{fign5m5t3} shows that the condition is sufficient also.
For the case of $k=3$, we have $n=7, \lambda=15$.
The graph in Figure
\ref{fign7m5t3} shows that the condition is sufficient also.

For $k=2r$ ($r\ge 2$), the condition
becomes $n=(m-t)k+1=4r+1$ and $\lambda =km=10r$. 
If we glue the $v_1$ node of $r$ copies of  $G_{5,1}$, we
get a $(t+1)$-color connected colored graph $G$ with
$n=4r+1$ and $\lambda =10r$. Thus the condition for the case
of $k=2r$ holds.

For $k=2r+1$ ($r\ge 2$), the condition becomes
$n=(m-t)k+1=4r+3$ and $\lambda=km=10r+5$.
If we glue glue the $v_1$ node of $r-1$ copies of 
$G_{5,1}$ and one copy of $G_{5,2}$, we get a $(t+1)$-color connected 
colored graph $G$ with $n=4(r-1)+1+6=4r+3$ and $\lambda =10(r-1)+15=10r+5$. 
Thus the condition for the case of $k=2r+1$ holds. 
This completes the induction.

For the condition 
``$n=(m-t)k+2$ and $\lambda \ge mk+t+1$'', we 
have $n=2k+2$ and $\lambda \ge 5k+4$.
We can add one 
node to the graph for the case ``$n=(m-t)k+1$ and $\lambda \ge km$''
 with $4$ edges (with distinct colors) 
to any four nodes. The resulting graph meets the requirements.
\hfill{Q.E.D.}

\vskip 5pt
\noindent
{\bf Open Questions:} We showed in this section that 
the conditions in Theorems \ref{dogthm} and \ref{necessarybound}
are sufficient for practical cases. It would be interesting 
to show that these conditions are also sufficient for general cases.
We leave this as an open question.

\section{Hardness results}
\label{hardsec}
We have given necessary and sufficient conditions for 
$(t+1)$-color connected colored edge graphs. Sometimes,
it is also important to determine whether a given graph is 
$(t+1)$-color connected. Unfortunately,
the following Theorem shows that the problem ceConnect is 
{\bf coNP}-complete. The ceConnect problem is defined as follows.

\vskip 5pt
\noindent 
INSTANCE: A colored edge graph $G=G(V,E,C,f)$, two nodes $A,B\in V$,
and a positive integer $t\le |C|$.

\noindent
QUESTION: Are $A$ and $B$ $t$-color connected?

Before we prove the hardness result, we first introduce 
the concept of color separator. For a colored edge graph 
$G=G(V,E,C,f)$, a color separator for two nodes $A$ and $B$ of 
the graph $G$ is a color set $C'\subseteq C$ such that the removal 
of all edges with colors in $C'$ from the graph $G$ will disconnect
$A$ and $B$. It is straightforward to observe that 
$A$ and $B$ are $(t+1)$-color connected if and only there is 
no $t$-size color separator for $A$ and $B$.

\begin{theorem}
\label{npcomplete}
The problem ceConnect  is {\bf coNP}-complete.
\end{theorem}

\noindent
{\bf Proof.} 
It is straightforward to show
that the problem is in {\bf coNP}. Thus it is sufficient to show that
it is {\bf NP}-hard. The reduction is from the Vertex Cover 
problem. The VC problem is as follows (definition taken from \cite{garey}):

\vskip 5pt
\noindent 
INSTANCE: A graph $G=(V,E)$ and a positive integer $t\le |V|$.

\noindent
QUESTION: Is there a vertex cover of size $t$ or less for $G$, 
that is, a subset $V'\subseteq V$ such that $|V'|\le t$ and, 
for each edge $(u,v)\in E$, at least one of $u$ and $v$ belongs to $V'$?

\vskip 5pt
\noindent
For a given instance $G=(V,E)$ of VC, we construct a 
colored edge graph $G_c=(V_c, E_c, f, C)$ as follows. First
assume that the vertex set $V$ is ordered as in $V=\{v_1,\ldots, v_n\}$. Let
$$\begin{array}{lll}
V_c&= &\left\{A,B\right\}\bigcup 
\left\{e_{(v_i,v_j)}:
(v_i,v_j)\in E\mbox{ and }i<j\right\}\\
E_c &= & \left\{(A, e_{(v_i,v_j)}), (e_{(v_i,v_j)}, B)
: (v_i,v_j)\in E\right\} \\
C&=&\left\{c_v: v\in V\right\}\\
f &=&\left\{
f(A, e_{(v_i,v_j)})=c_{v_i}, f(e_{(v_i,v_j)}, B)=c_{v_j}: 
(v_i,v_j)\in E, i<j\right\} 
\end{array}$$
In the following, we show that there is a vertex cover of size $t$ 
in $G$ if and only if there is a $t$-color edge separator for $G_c$.

Without loss of generality, assume that $V'=\{v'_1, \ldots, v'_k\}$ is 
a vertex cover for $G$. Then it is straightforward to show that
$C'=\{c_{v'_i}: v'_i\in V'\}$ is a color separator for $G_c$ since 
each incoming path for $B$ in $G_c$ contains two colors 
corresponding to one edge $(v_i,v_j)$ in $G$.

For the other direction, assume that 
$C'=\{c_{v'_i}: i=1, \ldots, t\}$ is a $t$-color 
separator for $G_c$. Let $V'=\{v'_i: c_{v'_i}\in C'\}$. 
By the fact that $C'$ is a color separator for $G_c$, for each 
edge $(v_i,v_j)\in E$ in $G$, 
the path $(A, e_{(v_i,v_j)}, B)$ in $G_c$ 
contains at least one color from $C'$. Since this path contains only two colors
$c_{v_i}$ and $c_{v_j}$, we know that $v_i$ or $v_j$ or both belong to 
$V'$. In another word, $V'$ is a $t$-size vertex cover for $G$.
This completes the proof of the Theorem.
\hfill{Q.E.D.}

\subsection*{Acknowledgement}
The first author of this paper would like to thank Prof. Doug Stinson and 
Prof. Ruizhong Wei for some discussions on generalized cover-free families.

\end{document}